\documentclass[a4paper,11pt]{article}
\usepackage{jcappub}

\usepackage{mathrsfs}     
\usepackage{empheq}       
\usepackage{mathtools}    
\usepackage{bm}           
\usepackage{dcolumn}      
\usepackage{cleveref}     
\usepackage{soul}         
\usepackage{comment}      
\usepackage{fontawesome5} 
\usepackage{orcidlink}    

\definecolor{darkblue}{rgb}{0.0,0.0,0.5}
\hypersetup{
    linkcolor=darkblue,
    citecolor=blue,
    urlcolor=darkblue,
}

\newcommand{\be}{\begin{equation}}
\newcommand{\ee}{\end{equation}}
\newcommand{\bs}{\begin{split}}
\newcommand{\bea}{\begin{eqnarray}}
\newcommand{\eea}{\end{eqnarray}}

\newcommand{\alb}{\alpha_B}
\newcommand{\alk}{\alpha_K}
\newcommand{\dl}{\delta}

\newcommand{\eps}{\epsilon}

\newcommand{\om}{\Omega_m}
\newcommand{\ode}{\Omega_{\rm de}}
\newcommand{\odet}{\Omega_{{\rm de},0}}
\newcommand{\wde}{w_{\rm de}}
\newcommand{\wtot}{w_{\rm tot}}

\newcommand{\rde}{\rho_{\rm de}}
\newcommand{\pde}{P_{\rm de}}

\newcommand{\gm}{G_{\rm matter}}
\newcommand{\gl}{G_{\rm light}}
\newcommand{\geff}{G_{\rm eff}}

\newcommand{\mpl}{M_{\rm Pl}^2}
\newcommand{\alm}{\alpha_M}
\newcommand{\kp}{\kappa}
\newcommand{\lam}{\lambda}

\newcommand{\fe}{f_{\rm early}}

\newcommand{\github}[1]{\href{#1}{\faGithub}}

\title{Charging Across the Phantom Divide with Modified Gravity}

\author[a]{Rodrigo Calder\'on\orcidlink{0000-0002-8215-7292}}
\author[b]{Eric V.\ Linder\orcidlink{0000-0001-5536-9241}}

\affiliation[a]{CEICO, Institute of Physics of the Czech Academy of Sciences,
  Prague, Czech Republic}
\affiliation[b]{Berkeley Center for Cosmological Physics \& Berkeley Lab,
  University of California, Berkeley, CA 94720, USA}

\emailAdd{calderon@fzu.cz}
\emailAdd{evlinder@lbl.gov}

\abstract{Cosmology where the effective dark energy crosses $w=-1$
can be realized in
Horndeski gravity with shift symmetric terms plus a linear potential.
We highlight the special role
of the nearly conserved scalar charge. The
theory is highly predictive for the early phantom behavior and
we identify three ways to cross $w=-1$. None of them recreate
conditions indicated by current data very well. The major
lesson is that such modified gravity with a potential lacking a cosmological constant
and only crossing $w=-1$ once (hence the less elaborate
models) has difficulty fitting current data.
We provide an interactive online application solving the system of evolution equations, for the reader to explore various scenarios at will.}

\begin{document}

\maketitle

\section{Introduction}

Current cosmological data indicate a beautifully bizarre
picture of cosmic expansion where the effective dark energy
equation of state crosses $w=-1$ \cite{DESI-Adame2024,DESI-calderon2024,DESI-lodha2025}. Such behavior does not
occur in the usual theories, and requires some
fundamentally new physics. Crossing $w=-1$ (the ``phantom divide'')
via coupling to the matter sector, either through
interaction with the dark energy or through nonminimal
gravitational coupling, tends to run into issues with
growth of large scale structure, including the cosmic
microwave background (CMB) integrated Sachs-Wolfe
effect and gravitational lensing \cite{2506.02122,2407.02558,DESI-MG:2024,2409.17019,2605.12415}.
Modified gravity
that does not change the matter coupling, however, would not
automatically have such issues, and is ripe for further
investigation (see e.g.\ \cite{tsuji,2509.17586,2512.13691,2512.03139}).
Ideas beyond the usual scalar-tensor gravity include
\cite{2505.24732,2507.00986,2508.01378,2503.22515}. Here however we work strictly within Horndeski gravity
with minimal matter coupling and a speed of gravitational wave propagation equal to the speed of light.

In \cite{2512.03139} we demonstrated how combining two
key observational constraints -- indication of an effective
dark energy equation of state crossing the phantom divide,
and the smallness of
any deviation in the gravitational coupling strength
(effective Newton's constant)
from general relativity (GR) -- implied
specific conditions on modified gravity theories
that do not change the matter coupling, e.g.\ shift
symmetric Horndeski gravity.

Here we investigate this in greater detail, factoring
in the interplay between the kinetic and gravitational
terms in the Horndeski action, conservation of scalar
charge, and the stability and no ghost conditions.
We focus on a cosmology where dark energy dominates
the expansion at late times and does not
recross the phantom divide; thus at late times $w>-1$.
To cross $w=-1$ a linear potential (like shift symmetric terms also somewhat protected against quantum corrections) is included.
Taking all the constraints into account,
we then numerically compute the cosmic evolution for
a realization of the theory.

\Cref{sec:model} reviews the modified gravity theory and
the cosmological evolution equations.
In~\Cref{sec:kin} we examine the characteristic kinetic and gravitational
structure of the modified gravity, and their interplay, requisite for
key observational properties and a healthy theory.
The theory is highly predictive for the early time behavior.  \Cref{sec:evo} derives the evolution equations in a form
suitable for numerical computation, which clarifies the
mechanisms for crossing $w=-1$.
We discuss the results in
\Cref{sec:res}, and three generalizations of the model in \Cref{sec:var},
including more general potentials, then conclude in \Cref{sec:concl}.

\section{Shift Symmetry and Charge} \label{sec:model}

Shift symmetry reins in quantum effects such as radiative
corrections and is a useful property to consider. However,
a purely shift symmetric Horndeski gravity cannot cross
$w=-1$ from $w<-1$ at early times to $w>-1$ at late times \cite{tsuji,2509.17586,2103.11195,1008.0048,0407107}, but requires a ``push'' from
a potential term. The potential itself will be subject to
quantum corrections though, unless it has some symmetry
properties. The simplest form is a linear potential
$V=\lam\phi$, which is protected against quantum loop corrections \footnote{The underlying reason is that $V=\lam\phi$ breaks the shift
symmetry $\phi\to\phi+c$ only softly: it shifts the Lagrangian by the constant $\lam c$ and so leaves the equations of motion unchanged. Since $V''=0$, no scalar mass is generated at
tree level, and corrections to $V$ must be proportional to the symmetry-breaking parameter $\lam$ itself, vanishing as $\lam\to0$; the potential is therefore technically natural in
the 't~Hooft sense.}  (see e.g.\ \cite{Kolda:1998wq,Peccei:2000rz,Doran:2002bc}).

Under our restrictions of minimal matter coupling and the speed of gravitational
wave propagation being the speed of light,\footnote{GW170817 constrains $c_T$ at
LIGO/Virgo frequencies, which lie near a natural cutoff scale for dark energy EFTs,
$\Lambda_3=(M_{\rm Pl}H_0^2)^{1/3}\sim260$ Hz, so
the bound need not extrapolate to cosmological scales \cite{deRham:2018red}. We nonetheless work under the
assumption $c_T=c$.} this brings the
Horndeski gravity Lagrangian to
\be
{\mathcal L}=\frac{1}{2}\,R+ K(\phi,X) -G_3(X)\,\Box\phi\ ,
\ee
with $K(\phi,X)\equiv\kp(X)-\lam\phi$, and
the Planck mass normalized so $M_{\rm Pl}^2=1$.
Note that $c_T=c$ alone would still permit a nonminimal coupling $G_4(\phi)$. We
further fix $G_4=1/2$, constant, so that the effective Planck mass does not run and $\alm=0$;
the absence of gravitational slip found in \Cref{sec:res} follows from this restriction. The
theory is to be understood as a low energy effective description of dark energy, applied here
in matter domination and later. Rather than a fixed cutoff, the relevant measure of validity is
the dimensionless kineticity and braiding $\alk$, $\alb$ (\Cref{sec:kin}), which we find $\ll1$
throughout, vanishing as $\ode\to0$ at early times.

The modified Friedmann equations and field equation of
motion are
\bea
3H^2&=&\rho_m+\rde\ ,\\
-2\dot H&=&\rho_m+P_m+\rde+\pde\ ,\\
0&=&\ddot\phi\,\left[K_X+2XK_{XX}+6H\dot\phi g_X\right]\notag\\
&\qquad&\quad+3H\dot\phi K_X+\lam+6g\left(\dot H+3H^2\right)\ , \label{eq:fd}
\eea
where we write $g\equiv XG_{3X}(X)$, and
a subscript $X$ denotes $d/dX$.
The effective dark energy density and pressure are
\bea
\rde&=&-K+2XK_X+6H\dot\phi g\ ,\label{eq:rho}\\
\pde&=&K-2g\ddot\phi\ . \label{eq:pde}
\eea
Neglecting radiation, the material pressure $P_m=0$.

We will also make use of a
special characteristic, the Noether charge density.
In shift symmetric
Horndeski theories the charge is conserved, but here it
still plays a useful role.
The scalar field equation (\ref{eq:fd}) takes the form
\be
\dot C+3HC=a^{-3}\,(a^3 C)\,\dot{}=-\lam\ , \label{eq:cdotlam}
\ee
with a convenient closed form solution for the charge density
\be
C\equiv\dot\phi K_X+6Hg=C_0 a^{-3}-\lam a^{-3}\int_0^a dA\,\frac{A^2}{H}\ .
\ee

In the purely shift symmetric case when $\lam=0$ then
we see $(a^3C)\,\dot{}=0$ and indeed the charge is
conserved. However, for the full solution we note that
since $1/H\sim a^{>0}$ is small at early times then the
second, integral term is small relative to the first
term. Thus, despite the presence of the potential, the
theory acts shift symmetric at early times. This is
important, since quantum corrections would be largest at
high energies, i.e.\ early times. This valuable property
does not hold for a general potential but is due to
adopting a linear potential, so the right hand side of
Eq.~(\ref{eq:cdotlam}) is constant.
We discuss other potentials in \Cref{sec:var}.

\section{Kinetic and Gravitational Structure} \label{sec:kin}

To achieve the observationally implied properties both of
phantom crossing and of limited deviations in the
gravitational strengths $\gm$ and $\gl$ (i.e.\ modified
Newton's constants in the modified Poisson equations of order Newton's constant)
from their general relativity values, the kinetic structure
of the Horndeski theory must take on specific characteristics,
as derived in \cite{2512.03139}.

Let us begin by defining the fractional contributions of
each term to the dark energy density:
\bea
f&\equiv&\frac{6H\dot\phi g}{\rde}\\
\eps&\equiv&\frac{\lam\phi}{\rde}\\
1-f-\eps&\equiv&\frac{-\kp+2XK_X}{\rde}=\frac{\kp(2n-1)}{\rde}\ , \label{eq:kinfrac}
\eea
where
\be
n(X)\equiv \frac{XK_X}{\kp}\ . \label{eq:ndef}
\ee
This was identified as a key quantity in \cite{2512.03139}.

At early times, as we have said above, $\eps\to0$.
Thus at early times $\rde\approx(1-f)\rde+f\rde$.
The limit where the kinetic term dominates is
$f\to0$, and the gravitational braiding term dominates as
$f\to1$, but we will in fact find that the two terms in the
charge density $C$ arising from the kinetic term
and the gravitational braiding $g$ term
must balance each other for a
successful theory. That is, they must each scale
similarly, so no one of them dominates, $0<\fe<1$.

\subsection{Early Time Limit and Balance} \label{sec:early}

At early times we find we cannot be dominated by either the
kinetic contribution from $\kp$ nor the braiding
contribution from $g$. The former will give an unstable
theory while the latter will give one with a ghost.
To show this, consider
the braiding and kineticity parameters \cite{bellsaw},
\bea
\alpha_B&=&\frac{2\dot\phi g}{H}\ ,\\
\alpha_K&=&\frac{2X}{H^2}\,(K_X+2XK_{XX}+6H\dot\phi g_X)\ .
\eea
Note the similarity of the terms in the
expression for $\alk$ to those in $\rde$. Indeed, we
have a progression of similar terms going from
the charge density to the dark energy density
(think of multiplying by $\dot\phi$) and then
to $\alk$ (think of taking a derivative with
respect to $X$). For example if one of the types
of terms dominates in one of these quantities,
it will dominate in all of them.

For the soundness of the theory,
we require the no ghost condition,
\be
\alpha\equiv\alk+\frac{3}{2}\,\alb^2\ge 0\ ,
\ee
and the stability condition, in terms
of non-negative sound speed squared of the scalar
perturbations,
\be
\alpha c_s^2=\left(1-\frac{\alb}{2}\right)\left(\alb-\frac{2\dot H}{H^2}\right)+\frac{\dot\alb}{H}-\frac{\rho_m+P_m}{H^2}\ge0\ .
\ee

First consider braiding domination at early times,
i.e.\ $f\to1$. Then
\be
\alk\to\ode\,\frac{6Xg_X}{g}\ ,
\ee
while $\alb\to\ode$. Since $\alb$ enters the no ghost
condition as the square, and $\ode\ll1$ at early times,
then the condition is simply $\alk\ge0$. Therefore we
require
\be
\frac{Xg_X}{g}=\frac{X}{X'}\frac{g'}{g}\ge0\ , \label{eq:gxreq}
\ee
where a prime denotes an e-fold
derivative, i.e.\ with respect to $\ln a$.
Since the $g$ term dominates the charge density then
$g\sim a^{-3}H^{-1}$ and
\be
\frac{g'}{g}\approx-\frac{3}{2}(1-\wtot)=-(2-q)\ ,
\ee
where $\wtot$ is the equation of state of the total
energy density, and $q$ is the cosmic deceleration
parameter.
The quantity $g'/g$ then must be negative for an early matter or radiation
dominated epoch. Furthermore,
\be
\rde\approx 6H\dot\phi g\approx \dot\phi C\sim \dot\phi a^{-3}\ .
\ee
We must have $\dot\phi$ growing with scale factor $a$,
and hence $X'>0$, if the dark energy density is ever to
overtake the matter density and give cosmic acceleration.
Thus, the requirement of Eq.~(\ref{eq:gxreq}) is violated and $\alk<0$ so braiding domination at early times
leads to a ghost.

The other extreme is kinetic domination at early times,
i.e.\ $f\to0$. Since this implies $\alb\to0$ then the
stability condition becomes
\be
\alpha c_s^2=\frac{-2\dot H}{H^2}-\frac{\rho_m+P_m}{H^2}=\frac{\rde+\pde}{H^2}\ge 0\ .
\ee
However, we require the dark energy to be phantom at
early times, and so the stability condition is violated.
Thus we cannot have kinetic domination at early times.

This is very interesting, and predictive. There must be
a balance between the two terms contributing to the charge
density. This early time balance implies the two terms
must scale the same, and so each one scales as $a^{-3}$.
This requirement will determine the early time form of
$\kp(X)$ and $g(X)$, and their relative amplitude.

\subsection{Early Time Limit and Equation of State} \label{sec:earlyeos}

The balance between the charge density terms requires
that each scales as $a^{-3}$, implying
\be
\dot\phi K_X\sim a^{-3}\ .
\ee
Denoting the asymptotic early time dependence as $\kp\sim X^n$, this gives
\be
X^{n-1/2}\sim a^{-3} \Rightarrow X\sim a^{-6/(2n-1)}\ . \label{eq:xearly}
\ee
Since the scaling between the charge density terms carries
through to dark energy density, $\rde\sim\kp\sim H\dot\phi g$,
then $\rde\sim X^n$. But the dark energy equation of state
is defined through the continuity equation as
\be
w\equiv -\frac{1}{3}\,\frac{d\ln\rde}{d\ln a}-1\quad\to\quad \frac{1}{2n-1}\ . \label{eq:wearly}
\ee
Thus at early times, asymptotically $w=1/(2n-1)$ is
constant. In order to have phantom behavior we require
$n_{\rm early}\equiv r\in[0,1/2]$. Note the predictivity: the kinetic power
law index $r$ must lie within a narrow range, and it
determines the dark energy equation of state, which is
asymptotically constant. Although the braiding term does
not explicitly enter into the equation of state, it must
exist in balance, i.e.\ this is not k-essence.

Note that the monomial form is less restrictive than it appears. The index $n$ was
defined in Eq.~(\ref{eq:ndef}) as the logarithmic slope $X\kp_X/\kp=d\ln\kp(X)/d\ln X$ of a general $\kp(X)$, not
as a monomial exponent, and $\kp$ and $g$ enter the evolution equations of
\Cref{sec:evo} only through this slope and $Xg_X/g=d\ln g(X)/d\ln X$. The monomial is simply the case of
constant slope; any $\kp(X)$ approaching the same asymptotic slope gives the same early time
behavior \footnote{up to the term $\beta=2Xn_X/n$ of Eq.~(\ref{eq:alkbeta}), which vanishes in the early-time 
limit.}.

The closer $r$ is to $1/2$, the more phantom the dark
energy is at early times (with $r=1/2$ representing a
phase transition from $w=-\infty$, $\rho_{\rm de}=0$).
If $r$ is close to 0, then the dark energy is only
mildly phantom at early times.

\subsection{Early Time Limit and Function Scaling} \label{sec:earlyfn}

We have seen
that the kinetic term asymptotically scales as $\kp(X)\sim X^r$.
Now for the gravitational $g$ term. The scaling of the
charge density term gives
\be
g\sim H^{-1}a^{-3}\sim H^{-1} X^{r-1/2}\ .
\ee
The early time asymptotic index for $g\sim X^b$ is
\be
m\equiv b_{\rm early}=\frac{(2-q)(2r-1)}{6}\to\frac{2r-1}{4}\ , \label{eq:gindexearly}
\ee
with the arrow giving the matter dominated
asymptote, where $q=1/2$.
This also guarantees that $\alb\ll1$ at early
times. Thus, in the early matter dominated epoch we have
\be
\kp(X)\sim X^r\, \qquad g(X)\sim X^{(2r-1)/4}\ ,
\ee
with $r\in[0,1/2]$.

\subsection{Early Time Limit and Amplitude} \label{sec:earlyamp}

To obtain the relative amplitude of the kinetic and
braiding terms we return to the no ghost and stability
conditions. Recall the fractional braiding contribution to the dark
energy density is $f$ and the fractional kinetic terms
contribution is $1-f$.

In the asymptotic early time limit,
\bea
\alb&=&f\ode\ ,\label{eq:albss}\\
\alk&=&\ode\,\left[6n(1-f)+f(2n-1)(2-q)\right]\ ,\label{eq:alkss} \\
\alpha c_s^2&=& [f+3(1+w)]\,\ode\left(1-\frac{f\ode}{2}\right)+(f\ode)'\notag\\
&\quad&-\frac{3}{2}f\ode(1-\ode)\\
&=&\frac{\ode}{2}\,\left[6(1+w)-f(1+6w)+2f'\right]\notag\\
&\quad&+\frac{f\ode^2}{2}\,(3w-f)\ .
\eea
Asymptotically $w=1/(2n-1)$
is constant, and $f'\to0$, so
we can analytically establish the conditions for
a healthy theory.

At early times $\alk\sim{\mathcal O}(\ode)$ dominates
over $\alb^2\sim{\mathcal O}(\ode^2)$, and so the no ghost
condition is essentially $\alk\ge0$. This becomes
\be
f_{\rm early}\le\frac{6n}{6n-(2n-1)(2-q)}=\frac{3(1+w)}{3(1+w)-(2-q)}<1\ ,
\ee
(where here $n$, $w$, $q$ are the early time asymptotic
values). As indicated above, since $1+w<0$ and $2-q>0$,
i.e.\ we have phantom dark energy in an early matter or
radiation dominated universe, then this implies $f<1$.

For stability, at early times the condition is
\be
f_{\rm early}\ge\frac{12n}{2n+5}=\frac{6+6w}{1+6w}>0\ ,
\ee
where again the phantom behavior requires $f>0$.
Putting these two together, the overall necessary (but not automatically sufficient) conditions for
a healthy theory are
\bea
0&<&\frac{12n}{2n+5}=\frac{6+6w}{1+6w}\le f_{\rm early} \label{eq:fearly}\\
&\quad&\le \frac{3(1+w)}{3(1+w)-(2-q)}=\frac{6n}{6n-(2n-1)(2-q)}<1\ . \notag
\eea
Thus the theory as written predicts not only the scaling of the
action terms but their relative amplitude in the early
time limit (practically, in matter domination at $z\approx1000$). As an example, if we take $r=1/4$, then
$w=-2$ and $6/11<f<2/3$ for the early matter dominated
epoch.

\section{Evolution Equations} \label{sec:evo}

As the field evolves -- recall from Eq.~(\ref{eq:xearly}) and
Eq.~(\ref{eq:wearly}) that $X\sim a^{-6w}$ early --
the linear potential $V=\lam\phi$ contribution to the
dark energy density increases and the shift symmetry is
broken. The field evolves away from its early time
asymptote. Here we present the full evolution equations
valid at all times.

\subsection{Analytic Form} \label{eq:evolanly}

First, note that the dark energy
equation of state,
\be
w=\frac{\pde}{\rde}=\frac{\kp-\lam\phi-6H\dot\phi g\,[\ddot\phi/(3H\dot\phi)]}{-\kp+\lam\phi+2XK_X+6H\dot\phi g}\ .
\ee
The quantity $\ddot\phi/(3H\dot\phi)=X'/(6X)$, as
can be seen by multiplying the left hand side numerator and denominator
by $\dot\phi$.

The field equation, Eq.~(\ref{eq:fd}), then becomes
\be
\alk\frac{X'}{X}=-6\ode\,\left[\frac{6n(1-f-\eps)}{2n-1}+f(2-q)+\eps\frac{\phi'}{\phi}\right]\ , \label{eq:fieldeqeps}
\ee
where $q$ is the cosmic deceleration parameter.
The property functions are
\bea
\alb&=&f\,\ode\ ,\\
\alk&=&6\ode\,\left[n(1-f-\eps)+f(Xg_X/g)\right]\ . \label{eq:alkX}
\eea
Note that since $f=f(X)$ and $n=n(X)$ then $\alb$ and $\alk$ are
{\it not\/} proportional to $\ode$ in general.

The dark energy equation of state is
\be
w=\frac{1-f-\eps}{2n-1}-\eps-f\,\frac{X'}{6X}\ .
\ee
One can easily verify that $w$ reduces to the previous Eq.~(\ref{eq:wearly}) at early times
when
$\eps\to0$, $X'/(6X)\to-1/(2n-1)$ from
Eq.~(\ref{eq:xearly}).
If in the late time asymptote, $\eps\to1$,
$f\to0$ then we have simply $w=-1$.
That is, the universe asymptotically
approaches a dark energy dominated de Sitter state,
and in that asymptotic future, $\alb\to0$,
$\alk\to0$ and we restore to GR. However, there are
many other ways to approach de Sitter and generally
$f$ (and $\alb$) stays finite and approaches a constant asymptotically.
This boundedness
may ameliorate strong gravitational deviations at the present and so
be more consistent with observations.

We still need to adopt the general forms for $\kp(X)$
and $g(X)$, though we know their asymptotic early behavior.
There is little physical guidance for this. We choose to minimize the number of free parameters by 
taking the asymptotic functional forms to hold for all times (for general functions presumably most any behavior could occur -- see Section~\ref{sec:var} for somewhat more complicated functions -- but current data has limited leverage). 
We take 
\bea
\kappa(X)&=&\kp_i\,(X/X_i)^r\quad\to\quad \kp_i\,(X/X_i)^{1/4}\\
g(X)&=&g_i\,(X/X_i)^b\quad\to\quad g_i\,(X/X_i)^{-1/8}\ ,
\eea
where we have chosen $r=1/4$ to fall midway in the
allowed range $r\in[0,1/2]$, and so $b=(2r-1)/4=-1/8$.
This also implies $Xg_X/g\equiv b=m=\,$ const.
The
amplitude of $g_i$ relative to $\kp_i$ is fixed by
$\fe$, and we can adopt $\fe=3/5$ to satisfy
Eq.~(\ref{eq:fearly}). (Note that for positive energy density
contributions $\kp_i<0$, $g_i>0$.)
We need to specify $\lam$, which we shall do through
$\eps$, and the various initial conditions; we also
have a constraint in $\odet=1-\om$.

\subsection{Numerical Form} \label{eq:evonum}

For numerical solution we convert
the Friedmann and scalar field equations
of motion to an autonomous dynamical system of coupled
ordinary differential
equations (see \cite{copeland,bahamonde} for details on such systems).
We also must specify the initial conditions.

Breaking the dark energy density into its dimensionless
components,
\bea
x_k&\equiv&\frac{\kappa}{3\mpl H^2}\\
x_g&\equiv&\frac{6H\dot\phi g}{3\mpl H^2}\\
x_\lam&\equiv&\frac{\lam\phi}{3\mpl H^2}\ .
\eea
Note that the $x_g=f\ode$, $x_\lam=\eps\ode$, etc.\ but
it is more convenient to use the $x$ variables.

The background expansion evolution has the logarithmic
derivative
\bea
\frac{d\ln H^2}{d\ln a}&=&-3(1+\wtot)\\
\wtot&=&\frac{\Omega_r}{3}+\wde\ode\ ,
\eea
while the energy density of component $i$ (matter, radiation,
dark energy) evolves (using the continuity equation) as
\be
\Omega'_i=-3\Omega_i\,(w_i-\wtot)\ . \label{eq:omiprime}
\ee 
We include radiation in the equations here for completeness, though we focus on the matter dominated and later eras. 

To evaluate the dark energy and total equations of
state it is convenient to define logarithmic derivatives
\bea
x_f&\equiv&\frac{\phi'}{\phi}\\
x_\phi&\equiv&\frac{\ddot\phi}{3H\dot\phi}=\frac{X'}{6X}\ ,
\eea
which will generally be of order one.
Thus we have
\bea
\ode&=&(2n-1)x_k+x_g+x_\lam \label{eq:odeode}\\
\wde&=&\frac{x_k-x_g x_\phi-x_\lam}{(2n-1)x_k+x_g+x_\lam} \label{eq:wnum}\\
\wde\ode&=&x_k-x_g x_\phi-x_\lam\ . \label{eq:wodenum}
\eea

The system of evolution equations becomes
\bea
x_k'&=&x_k\,\left[3(1+\wtot)+6n x_\phi\right] \label{eq:xkevo}\\
x_g'&=&x_g\,\left[\frac{3}{2}(1+\wtot)+3x_\phi\left(1+\frac{2Xg_X}{g}\right)\right] \label{eq:xgevo}\\
x_\lam'&=&x_\lam\,\left[3(1+\wtot)+px_f\right]\ , \label{eq:xlamp}
\eea
where $p\equiv d\ln V/d\ln\phi=1$ for the linear potential.
The field \Cref{eq:fieldeqeps} looks like
\be
x_\phi=\frac{-1}{\alk}\,\left[6nx_k+\frac{3}{2}\,(1-\wtot)\,x_g+px_\lam x_f\right]\ .  \label{eq:fdeqcon}
\ee
Since $\wtot$ involves $x_\phi$, one can instead substitute
in the expression for $\wtot$, bring the $x_\phi$ terms to
the same side, and use the expression
\bea
x_\phi&=&\frac{-1}{\alk+(3/2)x_g^2}\,\left[x_k\left(6n-\frac{3x_g}{2}\right)+\frac{3x_g}{2}\left(1-\frac{\Omega_r}{3}\right)\right.\notag\\
&\quad&\left.+x_\lam\left(px_f+\frac{3x_g}{2}\right)\right]\ . \label{eq:xphialg}
\eea
The property functions, suitable for testing the no ghost
and stability conditions, are
\bea
\alb&=&x_g\\
\alk&=&6n(2n-1)x_k+6x_g\,\frac{Xg_X}{g}\ . \label{eq:alkode}
\eea
Finally,
\be
x_f'=x_f\,\left[3x_\phi+\frac{3}{2}\,(1+\wtot)-x_f\right]\ . \label{eq:dxf}
\ee
All these equations reduce
properly to the quintessence limit, i.e.\ $\kp=X$, $g=0$.

For our present purposes, $n$ ($=r$) and $m\equiv Xg_X/g$ are simply numbers,
e.g.\ $1/4$ and $-1/8$. Note that the value $-1/8$ comes
from assuming we start evolution in the matter dominated era in Eq.~(\ref{eq:gindexearly}).
In the $x_\lam'$
equation, note that $x_f$ is more generally multiplied by $d\ln V/d\ln\phi$, but this is 1 for the linear potential.
We treat this, and $n(a)$, $m(a)$, in
\Cref{sec:var}. Initial conditions are discussed in the Appendix~\ref{sec:ic}.

\section{Results} \label{sec:res}

We present numerical results for the dark energy equation
of state $w(z)$, the property functions $\alb(z)$ and
$\alk(z)$, and other variables of interest.

\begin{figure}
    \centering
    \includegraphics[width=\textwidth]{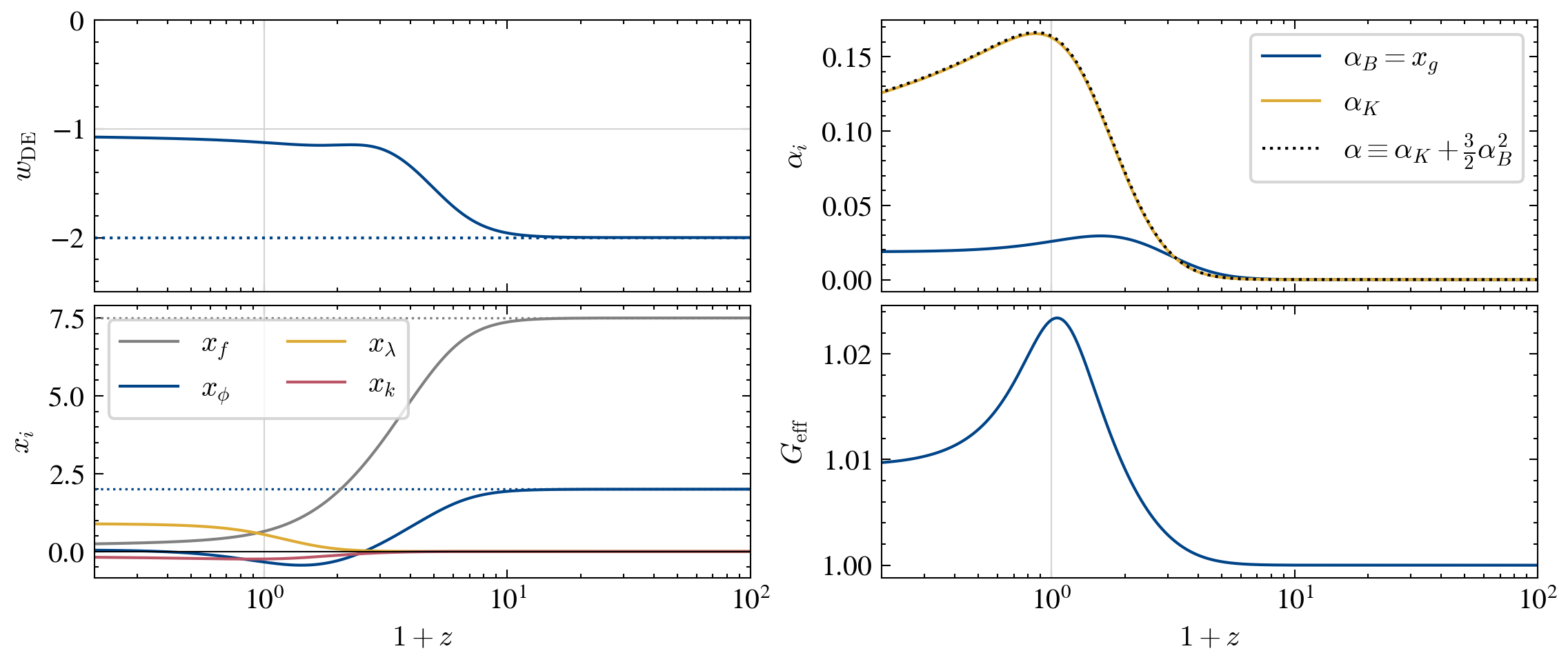}
    \caption{The fiducial case with a linear potential and monomials in $\kappa(X)$ and $G_3(X)$ fails to cross $w=-1$, but is otherwise well behaved. The fractional dark energy density today $\odet=0.69$. [Top panel] The effective dark energy equation of state smoothly evolves from a value constant in the past (due to charge conservation) to a de Sitter future. [Second panel] The property functions have modest deviations from their general relativity values of zero, and the no ghost condition $\alk+(3/2)\alb^2\ge0$ is obeyed. [Third panel] The dark energy density components and field evolution follow the charge conserved behaviors at early times before evolving. [Bottom panel] The effective gravitational coupling (same for matter and light) has only modest deviation from the general relativity value of one.}
    \label{fig:fidevo}
\end{figure}

Figure~\ref{fig:fidevo} shows these for our fiducial case where
$n=1/4$, $m=-1/8$, $f_{\rm early}=0.6$. The numerical
solutions at early times follow the analytic behaviors
derived in Section~\ref{sec:kin}. However, this case --
or any variation within the bounds of the analysis,
i.e.\ $n\in[0,1/2]$, and then $m$ and $f_{\rm early}$ given by
Eqs.~(\ref{eq:gindexearly}) and (\ref{eq:fearly}) --
does not achieve phantom crossing. We address the
necessary variations in Section~\ref{sec:var}, but
first let us examine the behaviors of the various functions.

The dark energy equation of state $w(z)$ has a long period
at high redshift where it evolves along the charge conservation
track, i.e.\ $w=1/(2n-1)$, until $x_\phi$ and $x_f$ begin
to drop. The equation of state then heads toward $-1$,
with the necessary condition for $w<-1$ (see next section) $x_\phi<1$ being achieved, but
this is not sufficient. The kinetic contribution $x_k$
increases rapidly in (negative) amplitude and this acts to
pull $w(z)$ phantom. However, the potential
contribution $x_\lam$ is also rapidly increasing and this
eventually brings $w(z)\to -1$. Although some cases have
$x_\phi<0$, i.e.\ $X$ stops increasing and starts decreasing,
we have $x_f>0$ so the field does not reverse its direction
of travel. Indeed, in the asymptotic future $|x_\phi|\ll1$
and the field almost freezes (from Eq.~\ref{eq:phiint} it
will slow to a growth $\sim\ln a$). Since $\alb=x_g$, then initially
$\alb$ increases, but it hits a ceiling as $x_k$ and $x_\lam$
dominate, decreasing $x_g$ and so $\alb$ at later times.

Apart from the expansion history, i.e.\ $w(z)$ just discussed, note also that we find
$\alb\ll1$ quite generally.
This greatly ameliorates the impact on
growth of large scale structure and on gravitational
lensing as well. Since there is no nonminimal coupling to
matter, i.e.\ a modified $G_4$ Horndeski term, then
$\alm=0$ and there is no gravitational slip:
$\geff=\gm=\gl$.
The two metric potentials are equal and the
gravitational coupling strengths entering the modified
Poisson equations for matter and relativistic particles
are equal (though not necessarily equal to Newton's
constant). We have
\be
\geff=1+\frac{\alb^2}{(2-\alb)\alb+2\alb'}\ ,
\ee
so a small and slowly varying $\alb$ does not give a
large change in growth of structure relative to general
relativity.
At early times, $\geff\to1$, then as $\alb$ becomes
more positive $\geff>1$, and then at late times
$\geff-1\to {\mathcal O}(\alb)\ll1$.

\section{More Complicated Functions} \label{sec:var}

The main focus
is to have a nearly shift symmetric modified gravity model
that crosses the phantom divide, ideally just near the present. By construction
the effective dark
energy starts phantom, with $w=1/(2n-1)\le-1$. Therefore
we need to understand how to get $w>-1$ at later times.

Note that from Eqs.~(\ref{eq:wnum}), (\ref{eq:wodenum}) the
condition $w>-1$ is
\be
2nx_k>x_g(x_\phi-1) \ . \label{eq:wgt}
\ee
Recall that $x_k<0$, $x_g>0$.
Once dark energy dominates,
$x_\phi$ tends to be close to zero, so this is mostly a
condition on $x_g/x_k$. To achieve $w>-1$ we need to keep $x_g$ sufficiently large relative to $x_k$. We find that for the fiducial case
of constant $n$, constant $m$, and $V=\lam\phi$ the
evolution smoothly approaches $w=-1$ without crossing
over.

Therefore, in this section we examine one at a time
the variation with scale factor (or $X$) of
$n$, $m$, and $V$. We expect that at late times we will
need to lower $n$, which also reduces the amplitude of
$x_k$ relative to $x_g$ by Eqs.~(\ref{eq:xkevo}) and
(\ref{eq:xgevo}), or lower $m$, which keeps $x_g$ high by
Eq.~(\ref{eq:xgevo}), or make $x_\phi$ more negative by
raising $x_\lam$, which comes from increasing the prefactor $p$
of $x_f$ in Eq.~(\ref{eq:xlamp}) by changing the potential
slope.

\begin{figure}
    \centering
    \includegraphics[width=\textwidth]{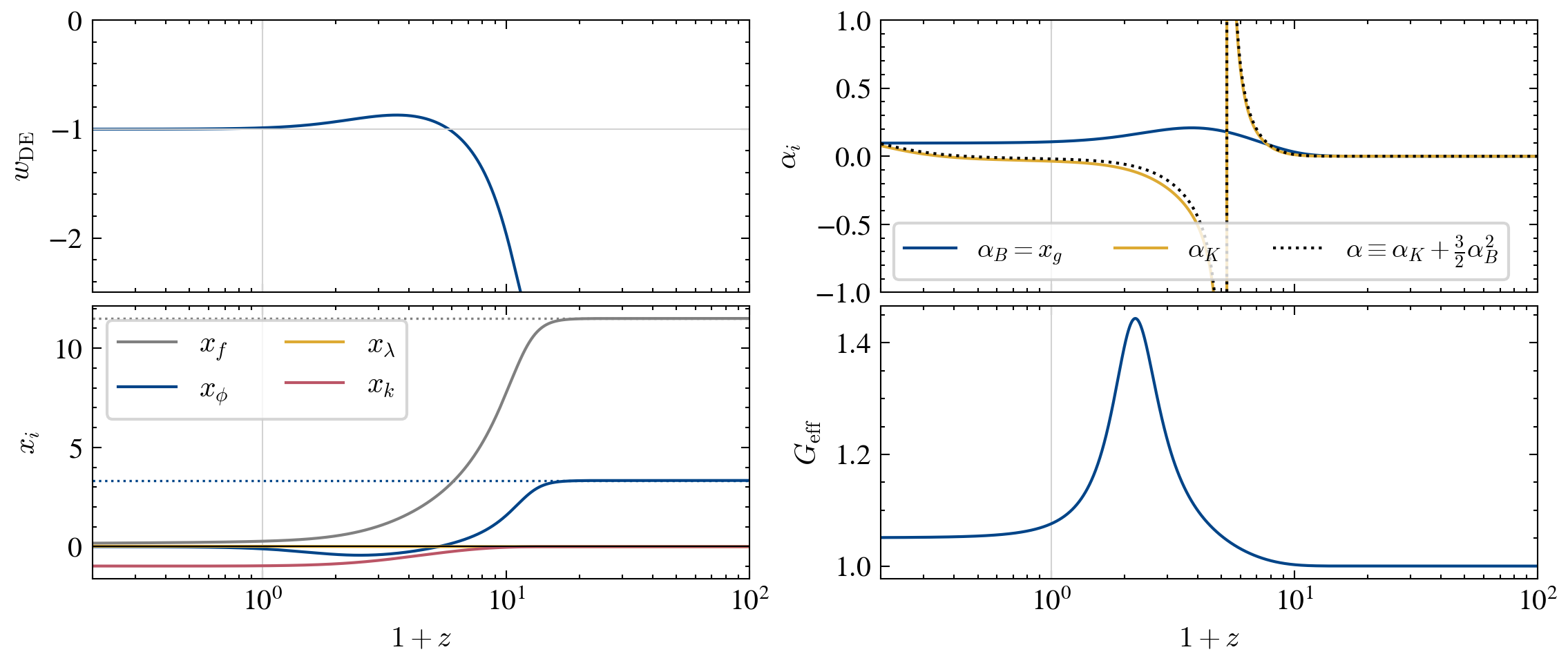}
    \caption{As Figure~\ref{fig:fidevo} but allowing $n$ to vary as $n=0.35-0.3\,\ode$ and $\odet\approx1$ to emphasize the future de Sitter behavior. The dark energy equation of state in the far past is $w_{\rm early}=-3.33$, following the charge conserved behavior. While now the model shows crossing of $w=-1$, the price is a divergence in $\alk$ (see \cref{sec:kx}) and so violation of the no ghost condition (and a larger $\geff$).}
    \label{fig:nvary}
\end{figure}

If we allow for more general $n(X)$, then the only explicit
change in the evolution equations is in
\be
\alk=6n(2n-1+\beta)x_k+6x_g\,\frac{Xg_X}{g}\ , \label{eq:alkbeta}
\ee
where $\beta\equiv 2Xn_X/n$. 
We approach this in two ways. 

\subsection{Gaining Intuition: Illustrative Parametrizations}

If we specify 
$n(X)$,
and hence a more complicated $K(X)$ than a monomial then we have both little physical guidance for the functional form -- beyond the early time asymptote of \Cref{sec:earlyeos} -- and additional parameters to fit for the data. Therefore we first gain
an illustrative sense of the impact by letting $n$ evolve with $a$,
specifically  $n=n_{\rm early}+\Delta n\,\Omega_{\rm de}$, where the only additional parameter is $\Delta n$. 
This form preserves all the early time behaviors from
Section~\ref{sec:kin} and smoothly transitions to a late time asymptote. 
Then
\be
\beta=\frac{\Delta n}{n(\Omega_{\rm de})}\,\frac{\Omega'_{\rm de}}{3x_\phi}=\frac{\Delta n}{n(\Omega_{\rm de})}\,\frac{-\Omega_{\rm de}(w_{\rm de}-w_{\rm tot})}{x_\phi}\ .
\ee
As $\beta$ (and hence $\alk$) involves $x_\phi$, both explicitly
and in $\wtot$, we have to rearrange Eq.~(\ref{eq:xphialg})
to give, after some algebra,
\bea
x_\phi&=&\frac{-N+6\Delta n(1-\ode)\,x_k(x_k-x_\lam)}{D+6\Delta n(1-\ode)\,x_kx_g}\\
N&\equiv&(6n-3x_g/2)x_k+(3x_g/2)(1-\Omega_r/3)\notag\\
&&+x_\lam(px_f+3x_g/2)\\
D&\equiv&6n(2n-1)x_k+6x_g\,(Xg_X/g)+3x_g^2/2\ .
\eea
One issue is that $\alk$ diverges at the point where $x_\phi=0$,
and hence violates the no ghost condition shortly afterward
(when $\alk\to -\infty$). An explicit
model for $K(X)$ can in fact avoid this issue, as we show in \Cref{sec:kx} below.
Nevertheless, it is useful to see how the
$x$ functions evolve to enable the $w=-1$ crossing.

Figure~\ref{fig:nvary} shows crossing of $w=-1$
for $\Delta n=-0.3$, with all other parameters held at their
test fiducial values ($n_{\rm early}=0.35$,
and preserving all the early time behaviors from
Section~\ref{sec:kin} so
$m=(2n_{\rm early}-1)/4=-0.075$, $f_{\rm early}=0.775$, with
$\Delta m=0$, $\log\Omega_{{\rm de},i}=-11$; note we make
no attempt to get a viable $\Omega_{{\rm de},0}$ as we want to see the approach to the de Sitter asymptote, and we use
the higher test value $n_{\rm early}=0.35$ to speed the initial
transition toward $w=-1$).

\begin{figure}
    \centering
    \includegraphics[width=\textwidth]{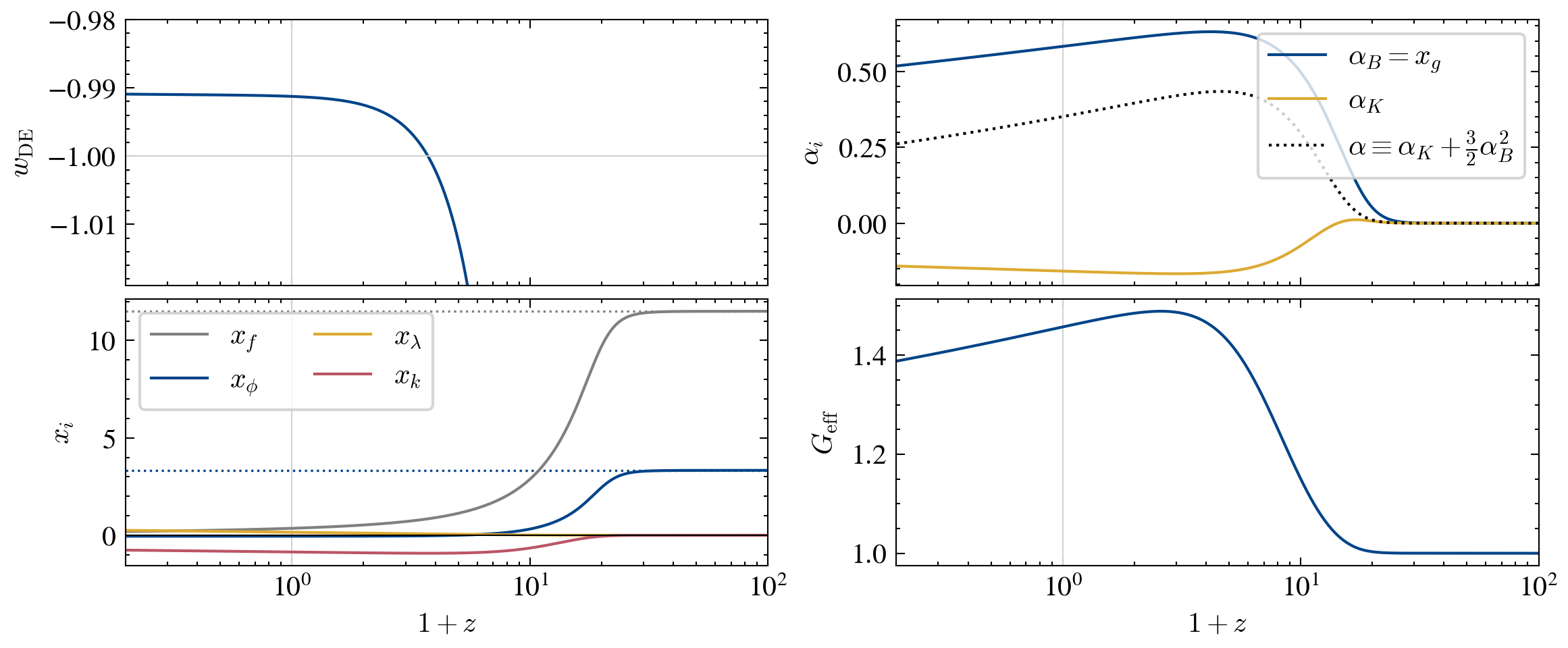}
    \caption{As Figure~\ref{fig:fidevo} but allowing $m$ to vary as $m=-0.075-0.125\,\ode$, and constant $n=0.35$ and $\odet\approx1$. The dark energy equation of state in the far past is $w_{\rm early}=-3.33$, following the charge conserved behavior, and it later crosses $w=-1$ but very mildly. There is no divergence in $\alk$ and the no ghost condition is satisfied, but there is a large deviation from general relativity.}
    \label{fig:mvary}
\end{figure}

Before we look at the case of a more complicated $K(X)$ in \Cref{sec:kx} below, let us consider varying the other parameters in the same way. 
Keeping $n$ constant (and the other test fiducial values) but now allowing
$m=m_{\rm early}+\Delta m\,\ode$, we see
this will only affect the $Xg_X/g=m$ term on the right hand side
of the $x_g'$ equation and in $\alk$. Note $\alk$ does
not diverge.
Figure~\ref{fig:mvary} shows crossing of $w=-1$ for
$\Delta m=-0.125$, although the crossing is exceedingly mild.

Finally, for more general $V(\phi)$, the evolution equations
are all the same when defining $x_\lam\equiv V/(3\mpl H^2)$,
with the exception that
now $p\ne1$.
That term arose from
\be
\frac{d\ln V}{d\ln a}=\frac{d\ln V}{d\ln\phi}\ \frac{d\ln\phi}{d\ln a}=p\,x_f\ .
\ee
Since $d\ln V/d\ln\phi=1$ in our fiducial case, we
had $p=1$, but now we consider varying it.
\begin{figure}
    \centering
    \includegraphics[width=\textwidth]{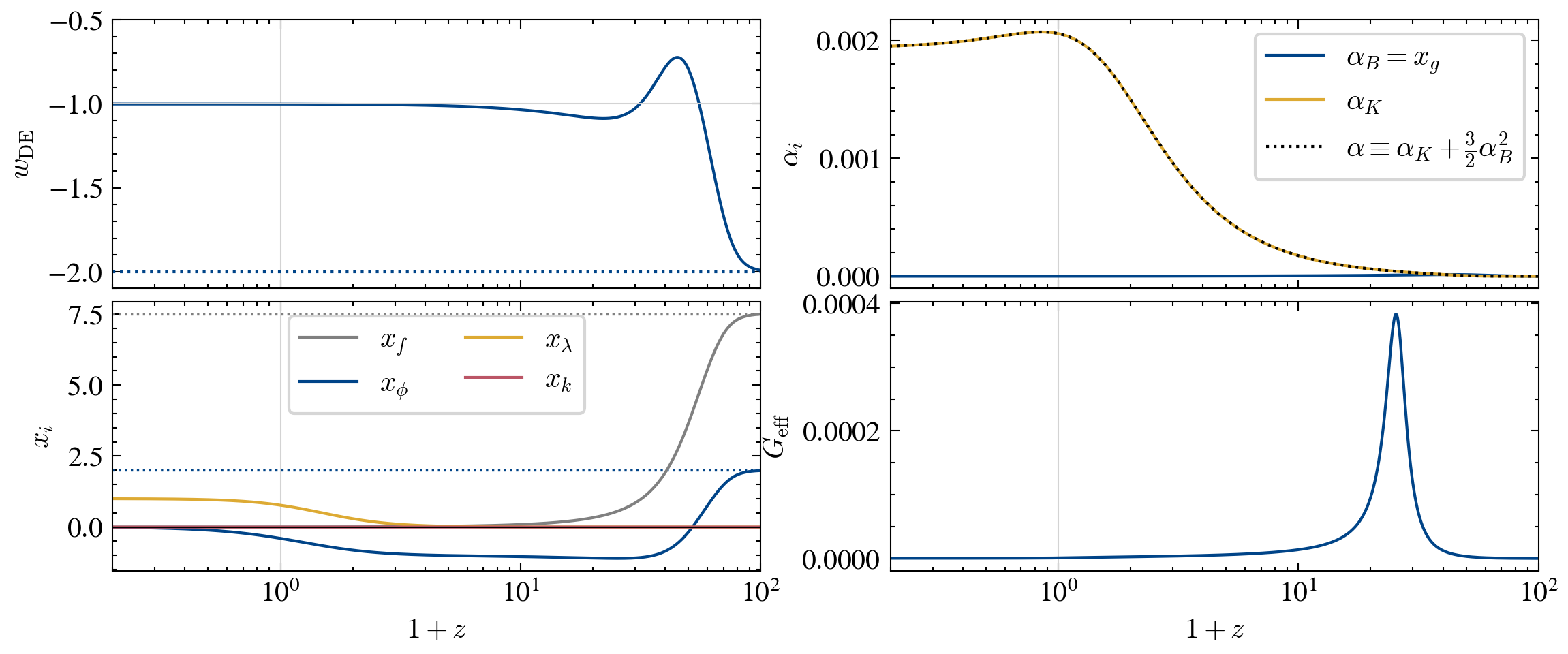}
    \caption{As Figure~\ref{fig:fidevo} but $p=2$ ($V\sim\phi^2$), $\fe=0.65$. The fractional dark energy density today is $\odet=0.69$. The dark energy density does cross $w=-1$ but much earlier than today, before crossing back and remaining in the phantom region. Deviations from general relativity are unobservably small.
    }
    \label{fig:vvary}
\end{figure}

To drive $x_\phi$ below one, and ideally negative, in order
to push $w$ towards, and hopefully above, $-1$, increasing
$d\ln V/d\ln\phi$ is helpful. This must be done at early
times, since at late times it merely moves the evolution
more quickly from the phantom to the de Sitter state. If one
employs $V\sim m^2\phi^2$, say, one can indeed achieve
$w>-1$ but the rapid evolution of $x_\phi$ means that
$w>-1$ at early times, and then $w$ drops back
to $w<-1$. If one uses $V\sim V_0+m^2\phi^2$, or a similar
model possessing an explicit cosmological constant, then
this does not increase $d\ln V/d\ln\phi$ at early times but rather makes it near zero. In each case, one opens 
oneself up to all the usual quantum corrections. Similarly
$V\sim e^{-\lam\phi}$ requires the field to start at
large $\phi$ rather than small $\phi$ as here, both raising
Planck scale issues and making $w$ at early times generally
not only $w>-1$ but often even 
$w>0$, potentially raising issues with fine tuning or domination over matter. 

Keeping $n$, $m$ constant we allow $p=p_{\rm early}+\Delta p\,\ode$ (our fiducial case of the linear
potential is $p_{\rm early}=1$, $\Delta p=0$).
Figure~\ref{fig:vvary} shows crossing of $w=-1$ for
$p_{\rm early}=2$, but $w$ quickly crosses back to $w<-1$
as $p$ becomes irrelevant due to $x_f$ approaching zero.

\subsection{Constructing an explicit, ghost-free $K(X)$} \label{sec:kx}

The $\alk$ divergence noted above for the $n(a)=n_{\rm early}+\Delta n\,\ode(a)$
prescription raises a natural question: can an explicit $K(X)$ realizing an analogous
variation be constructed, and is that divergence a generic feature of varying $n$, or an
artefact of prescribing the variation in the scale factor rather than in $X$? We show here
that the first has an affirmative answer, derive a closed-form family,
and show that for any such genuine $n(X)$ the combination $n\beta=2Xn_X$ that actually enters
$\alk$ is bounded, so the $\alk$ divergence at $x_\phi=0$ disappears. The ghost is therefore an
artefact of the scale factor parametrization used above, due as we will see to the vanishing of the Jacobian $dX/da$.

Recall from \eqref{eq:ndef} that $n(X)\equiv XK_X/\kp=d\ln\kp/d\ln X$ for a general $\kp(X)$;
no monomial is assumed. Written this way the definition is a first-order linear ODE for
$\ln\kp$ in the variable $u\equiv\ln X$, and integrates immediately:
\be
\kp(X)\;=\;\kp_0\,\exp\!\left[\int^{\ln X} n(u)\,du\right]. \label{eq:kappaquad}
\ee
Thus any prescribed $n(X)$ corresponds to an explicit $\kp(X)$ in general; what remains is to
choose an $n(X)$ and evaluate the integral. The same statement holds on the braiding side:
since $g\equiv XG_{3X}$ and the equations depend on $g$ only through $Xg_X/g=d\ln g/d\ln
X\equiv m(X)$, one has $g(X)=g_0\exp[\int m\,du]$, and then $G_3(X)=\int g\,dX/X$.

We want an $n(X)$ that preserves every early-time result of \Cref{sec:kin} --- which requires
$n\to n_{\rm early}$ as $X\to0$ --- and that moves smoothly to a different late-time value. The
natural choice is a sigmoid in $u=\ln X$,
\be
n(u)\;=\;n_{\rm early}+\Delta n\,S(u)\ ,\qquad
S(u)\;=\;\frac{1}{1+e^{-(u-u_0)/\dl}}\ , \label{eq:sigmoid}
\ee
with centre $u_0=\ln X_0$ and width $\dl>0$. Since $X\to0$ at early times for the phantom range
$n_{\rm early}\in[0,1/2]$, we have $S\to0$ and $n\to n_{\rm early}$ there, as required.
Evaluating \eqref{eq:kappaquad} with \eqref{eq:sigmoid}, and using
$\int S\,du=\dl\,\ln\!\left(1+e^{(u-u_0)/\dl}\right)$,
\be
\kp(X)\;=\;\kp_0\;X^{\,n_{\rm early}}\left[1+\left(\frac{X}{X_0}\right)^{1/\dl}\right]^{\Delta n\,\dl}. \label{eq:kappaclosed}
\ee
This is an explicit, elementary $K(X)=\kp(X)-\lam\phi$. One checks directly that
$d\ln\kp/d\ln X$ returns \eqref{eq:sigmoid}. Its
limits are the expected ones,
\be
\kp\to\kp_0X^{n_{\rm early}}\quad (X\ll X_0)\ ,\qquad
\kp\to\kp_0\,X_0^{-\Delta n}\,X^{\,n_{\rm early}+\Delta n}\quad (X\gg X_0)\ ,
\ee
so the model interpolates between two monomials with indices $n_{\rm early}$ and
$n_{\rm early}+\Delta n$. The monomial of \Cref{eq:evolanly} is recovered at $\Delta n=0$.

Recall that \eqref{eq:alkbeta} carries the varying-slope correction, with
$\beta\equiv2Xn_X/n$. For the model $n(X)$ of \eqref{eq:sigmoid}, $\beta$ follows from
$dn/du=\Delta n\,S(1-S)/\dl$:
\be
\beta(u)\;=\;\frac{2\,\Delta n\,S(u)\left[1-S(u)\right]}{n(u)\;\dl}\ ,
\qquad
|n\beta|\;\le\;\frac{|\Delta n|}{2\,\dl}\ , \label{eq:nbetabound}
\ee
using $S(1-S)\le1/4$. The product $n\beta$ (which is what enters in $\alk$) is smooth and
\emph{bounded} everywhere. In particular it contains no reference to $x_\phi$, so $\alk$ does
not diverge when the trajectory passes through $x_\phi=0$, and no ghost is triggered there.

Contrast the scale-factor prescription $n(a)=n_{\rm early}+\Delta n\,\ode(a)$ used above, for which
we found
\be
\beta=\frac{\Delta n}{n(\ode)}\frac{\ode'}{3x_\phi}\ . \label{eq:betaaX}
\ee
The origin of the $1/x_\phi$ is now transparent. Prescribing $n$ as a function of $a$ and then
asking for $n_X$ requires the chain rule $n_X=n'/X'$, and with $x_\phi\equiv X'/(6X)$ this
introduces exactly a factor $1/x_\phi$: it is the Jacobian of the map $a\mapsto X$. That map
degenerates precisely where $X$ turns around, i.e.\ at $x_\phi=0$, and there the
$a$-prescription ceases to correspond to \emph{any} $K(X)$. The divergence of $\alk$ is the
dynamical system reporting that failure, not a statement about the underlying theory. An
$X$-prescription, by contrast, remains well defined even when $X$ is non-monotonic in $a$,
which is a case of interest here -- some cases indeed have $x_\phi<0$, i.e.\ $X$ stops increasing and
starts decreasing.

\begin{figure}
    \centering
    \includegraphics[width=\linewidth]{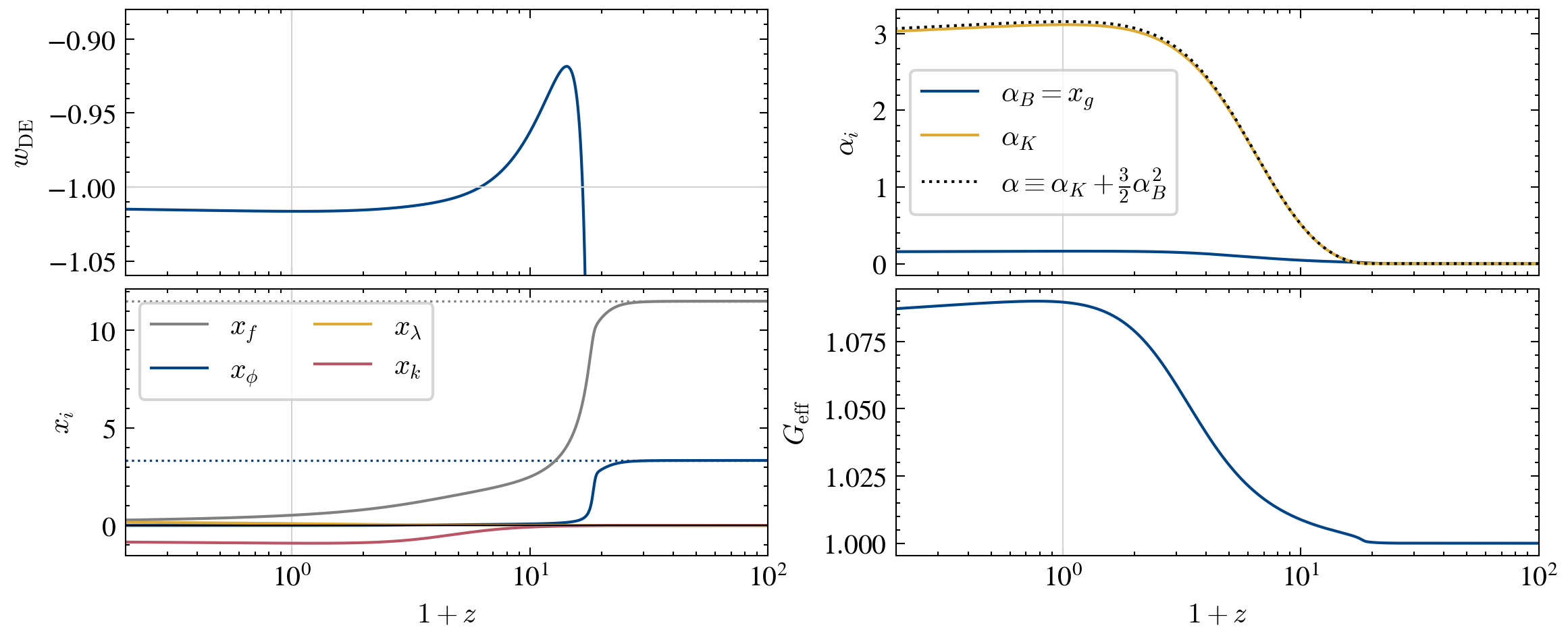}
    \caption{Using the $n(X)$ sigmoid prescription rather than the $n(a)$ approach of Fig.~\ref{fig:nvary}, still with $n_{\rm early}=0.35$, $\Delta n=-0.3$, gives a qualitatively similar $w(a)$ without any divergence leading to a ghost. However, it still does not look like our cosmology.}
    \label{fig:results_nX}
\end{figure}

Despite curing the $\alk$ divergence, the reconstructed family \eqref{eq:kappaclosed} fares no
better in producing a viable cosmology. 
Figure~\ref{fig:results_nX} shows the analog of Figure~\ref{fig:nvary}, having $n_{\rm early}=0.35$ and $\Delta n=-0.3$ but prescribing $n(X)$ rather than $n(a)$. 
This approach does indeed cross $w=-1$ and is ghost free, 
but does not look like our cosmology. 
Scanning over $n_{\rm early}$, $\Delta n$, and the now additional 
sigmoid parameters $u_0$, $\dl$ (together with $f_{\rm early}$ and $\Omega_{{\rm de},i}$), we
find no case with a crossing of $w=-1$ near the present, $\odet\simeq0.7$, and the no
ghost and stability conditions satisfied throughout. Curing the divergence therefore resolves
the theoretical objection but does not by itself yield a viable phantom-crossing cosmology.

Thus our approach emphasizing simplicity -- or at least parsimony (e.g.\ monomial functions of $X$ and linear function of $\phi$ in the potential), quantum protection,
and avoidance of an explicit cosmological constant does not
succeed in giving a viable cosmology.
We seem to have to live with some nonideality -- either giving
up quantum protection (e.g.\ early shift symmetry), adding a
cosmological constant (also quantum unprotected) or multiple phantom crossings, or adding
nonminimal coupling to matter -- if we want to cross the
phantom divide near the present as the data imply.

\section{Conclusions} \label{sec:concl}

Modified gravity has several free functions entering the
action that enable the effective dark energy equation of state
to cross the phantom divide from $w<-1$ to $w>-1$. We can
employ principles of protections against quantum radiative
correction, e.g.\ by symmetries, naturalness, and
observations indicating closeness to general relativity
to simplify the action.

For example we might seek to avoid nonminimal coupling
to the matter sector, inclusion of an explicit cosmological
constant, and possibly transPlanckian scalar field values.
As the most concise class of modified gravity -- shift
symmetric Horndeski gravity with no potential -- cannot cross
the phantom divide, we explored here the next best choice:
a linear potential, which is quantum protected.

Indeed, at early times when the contribution of the
potential is small, this acts like a shift symmetric theory
and the resulting charge conservation
is highly predictive for the balance between kinetic
and braiding terms, dark energy equation of
state, field evolution, and ghost free and stability
conditions. Alas we have shown that the linear potential
is not effective at crossing the phantom divide, at least for the power law (or near power law) functions of $X$ used.

Considering three elaborations on the model, we find
that more complicated kinetic or braiding terms can cross
$w=-1$, at least temporarily, but have problems
behaving as current data imply. A more complicated potential
(hence giving up the early time shift symmetry and predictivity)
can also cross $w=-1$, but again does not behave as desired --
unless one also allows large field values, initial equation
of state $w>-1$ (and sometimes $w>0$), or a cosmological
constant, i.e.\ lifted potential.

We are thus forced to introduce some complexity and
nonideality relative to our original goal, and these extra
terms bring with them additional uncertainty on how to
determine the appropriate functional forms, and additional
parameters. Entities have multiplied, but this seems driven
by necessity to fit current data with modified gravity.

To enable wider exploration by the community, we make public
an online interactive app \github{https://rcalderonb6.github.io/phantom-X/} solving the dynamical evolution system
for user specified parameter choices, available at
\url{https://rcalderonb6.github.io/phantom-X/}.

\section*{Acknowledgments}

R.C. is funded by the Czech Ministry of Education, Youth and Sports (MEYS) and European Structural and Investment Funds (ESIF) under project number CZ.02.01.01/00/22\_008/0004632.

\bibliographystyle{JHEP}
\bibliography{refs}

@Article{DESI-Adame2024,
  author        = {Adame, A.G. and others},
  collaboration = {DESI},
  title         = {{DESI 2024 VI: cosmological constraints from the measurements of baryon acoustic oscillations}},
  journal       = {JCAP},
  year          = {2025},
  volume        = {02},
  pages         = {021},
  eprint        = {2404.03002},
  archivePrefix = {arXiv},
}

@Article{DESI-calderon2024,
  author        = {Calderon, R. and others},
  collaboration = {DESI},
  title         = {{DESI 2024: reconstructing dark energy using crossing statistics with DESI DR1 BAO data}},
  journal       = {JCAP},
  year          = {2024},
  volume        = {10},
  pages         = {048},
  eprint        = {2405.04216},
  archivePrefix = {arXiv},
}

@Article{DESI-lodha2025,
  author        = {Lodha, K. and others},
  collaboration = {DESI},
  title         = {{Extended Dark Energy analysis using DESI DR2 BAO measurements}},
  journal       = {Phys. Rev. D},
  year          = {2025},
  volume        = {112},
  pages         = {083511},
  eprint        = {2503.14743},
  archivePrefix = {arXiv},
}

@Article{2506.02122,
  author        = {Linder, E.V.},
  title         = {{Uplifting, Depressing, and Tilting Dark Energy}},
  year          = {2025},
  eprint        = {2506.02122},
  archivePrefix = {arXiv},
}

@Article{2407.02558,
  author        = {Chudaykin, A. and Kunz, M.},
  title         = {{Modified gravity interpretation of the evolving dark energy in light of DESI data}},
  journal       = {Phys. Rev. D},
  year          = {2024},
  volume        = {110},
  pages         = {123524},
  eprint        = {2407.02558},
  archivePrefix = {arXiv},
}

@Article{DESI-MG:2024,
  author        = {Ishak, M. and others},
  collaboration = {DESI},
  title         = {{Modified gravity constraints from the full shape modeling of clustering measurements from DESI 2024}},
  journal       = {JCAP},
  year          = {2025},
  volume        = {09},
  pages         = {053},
  eprint        = {2411.12026},
  archivePrefix = {arXiv},
}

@Article{2409.17019,
  author        = {Wolf, W.J. and Ferreira, P.G. and {Garc\'{\i}a-Garc\'{\i}a}, C.},
  title         = {{Matching current observational constraints with nonminimally coupled dark energy}},
  journal       = {Phys. Rev. D},
  year          = {2025},
  volume        = {111},
  pages         = {L041303},
  eprint        = {2409.17019},
  archivePrefix = {arXiv},
}

@Article{2605.12415,
  author        = {Ye, G. and Chudaykin, A. and Bonvin, C. and Kunz, M.},
  title         = {{Late-time reconstruction of non-minimally coupled gravity with a smoothness prior}},
  year          = {2026},
  eprint        = {2605.12415},
  archivePrefix = {arXiv},
}

@Article{tsuji,
  author        = {Tsujikawa, S.},
  title         = {{Crossing the phantom divide in scalar-tensor and vector-tensor theories}},
  journal       = {Phys. Rev. D},
  year          = {2026},
  volume        = {113},
  pages         = {L041301},
  eprint        = {2508.17231},
  archivePrefix = {arXiv},
}

@Article{2509.17586,
  author        = {Wolf, W.J. and Ferreira, P.G. and {Garc\'{\i}a-Garc\'{\i}a}, C.},
  title         = {{Cosmological constraints on Galileon dark energy with broken shift symmetry}},
  journal       = {Phys. Rev. D},
  year          = {2026},
  volume        = {113},
  pages         = {023551},
  eprint        = {2509.17586},
  archivePrefix = {arXiv},
}

@Article{2512.13691,
  author        = {Cataneo, M. and Koyama, K.},
  title         = {{Non-parametric exploration of minimally coupled gravity with phantom crossing}},
  year          = {2025},
  eprint        = {2512.13691},
  archivePrefix = {arXiv},
}

@Article{2512.03139,
  author        = {Linder, E.V.},
  title         = {{Cosmology after Phantom Crossing by Horndeski Gravity}},
  year          = {2025},
  eprint        = {2512.03139},
  archivePrefix = {arXiv},
}

@Article{2505.24732,
  author        = {Cai, Y. and Ren, X. and Qiu, T. and Li, M. and Zhang, X.},
  title         = {{Quintom theory of dark energy after DESI DR2}},
  journal       = {Nat. Sci. Rev.},
  year          = {2026},
  pages         = {nwag115},
  eprint        = {2505.24732},
  archivePrefix = {arXiv},
}

@Article{2507.00986,
  author        = {Hell, A. and Sasaki, M.},
  title         = {{Accelerating Universe from Constraints}},
  year          = {2025},
  eprint        = {2507.00986},
  archivePrefix = {arXiv},
}

@Article{2508.01378,
  author        = {Yao, Z. and Ye, G. and Silvestri, A.},
  title         = {{General Model for Dark Energy Crossing the Phantom Divide}},
  journal       = {JCAP},
  year          = {2025},
  volume        = {10},
  pages         = {078},
  eprint        = {2508.01378},
  archivePrefix = {arXiv},
}

@Article{2503.22515,
  author        = {Ye, G. and Cai, Y.},
  title         = {{Null energy condition violation and beyond Horndeski physics in light of DESI DR2 data}},
  journal       = {Phys. Rev. D},
  year          = {2025},
  volume        = {112},
  pages         = {L121301},
  eprint        = {2503.22515},
  archivePrefix = {arXiv},
}

@Article{2103.11195,
  author        = {Traykova, D. and Bellini, E. and Ferreira, P.G. and {Garc\'{\i}a-Garc\'{\i}a}, C. and Noller, J. and {Zumalac\'arregui}, M.},
  title         = {{Theoretical priors in scalar-tensor cosmologies: Shift-symmetric Horndeski models}},
  journal       = {Phys. Rev. D},
  year          = {2021},
  volume        = {104},
  pages         = {083502},
  eprint        = {2103.11195},
  archivePrefix = {arXiv},
}

@Article{1008.0048,
  author        = {Deffayet, C. and Pujolas, O. and Sawicki, I. and Vikman, A.},
  title         = {{Imperfect Dark Energy from Kinetic Gravity Braiding}},
  journal       = {JCAP},
  year          = {2010},
  volume        = {10},
  pages         = {026},
  eprint        = {1008.0048},
  archivePrefix = {arXiv},
}

@Article{0407107,
  author        = {Vikman, A.},
  title         = {{Can dark energy evolve to the Phantom?}},
  journal       = {Phys. Rev. D},
  year          = {2005},
  volume        = {71},
  pages         = {023515},
  eprint        = {astro-ph/0407107},
  archivePrefix = {arXiv},
}

@Article{bellsaw,
  author        = {Bellini, E. and Sawicki, I.},
  title         = {{Maximal freedom at minimum cost: linear large-scale structure in general modifications of gravity}},
  journal       = {JCAP},
  year          = {2014},
  volume        = {07},
  pages         = {050},
  eprint        = {1404.3711},
  archivePrefix = {arXiv},
}

@Article{copeland,
  author        = {Copeland, E.J. and Sami, M. and Tsujikawa, S.},
  title         = {{Dynamics of dark energy}},
  journal       = {Int. J. Mod. Phys. D},
  year          = {2006},
  volume        = {15},
  pages         = {1753},
  eprint        = {hep-th/0603057},
  archivePrefix = {arXiv},
}

@Article{bahamonde,
  author        = {Bahamonde, S. and Boehmer, C. and Carloni, S. and Copeland, E.J. and Fang, W. and Tamanini, N.},
  title         = {{Dynamical systems applied to cosmology: Dark energy and modified gravity}},
  journal       = {Phys. Rept.},
  year          = {2018},
  volume        = {775},
  pages         = {1--122},
  eprint        = {1712.03107},
  archivePrefix = {arXiv},
}

@article{Kolda:1998wq,
    author = "Kolda, Christopher F. and Lyth, David H.",
    title = "{Quintessential difficulties}",
    eprint = "hep-ph/9811375",
    archivePrefix = "arXiv",
    reportNumber = "LBL-42536, LBNL-42536, LANCS-TH-9821",
    doi = "10.1016/S0370-2693(99)00657-7",
    journal = "Phys. Lett. B",
    volume = "458",
    pages = "197--201",
    year = "1999"
}

@inproceedings{Peccei:2000rz,
    author = "Peccei, R. D.",
    title = "{Light scalars in cosmology}",
    booktitle = "{4th International Symposium on Sources and Detection of Dark Matter in the Universe (DM 2000)}",
    eprint = "hep-ph/0009030",
    archivePrefix = "arXiv",
    pages = "98--104",
    month = "2",
    year = "2000"
}

@article{Doran:2002bc,
    author = "Doran, Michael and Jaeckel, Joerg",
    title = "{Loop corrections to scalar quintessence potentials}",
    eprint = "astro-ph/0203018",
    archivePrefix = "arXiv",
    reportNumber = "HD-THEP-02-08",
    doi = "10.1103/PhysRevD.66.043519",
    journal = "Phys. Rev. D",
    volume = "66",
    pages = "043519",
    year = "2002"
}

@article{deRham:2018red,
    author = "de Rham, Claudia and Melville, Scott",
    title = "{Gravitational Rainbows: LIGO and Dark Energy at its Cutoff}",
    eprint = "1806.09417",
    archivePrefix = "arXiv",
    primaryClass = "hep-th",
    doi = "10.1103/PhysRevLett.121.221101",
    journal = "Phys. Rev. Lett.",
    volume = "121",
    number = "22",
    pages = "221101",
    year = "2018"
}

\appendix

\section{Initial conditions} \label{sec:ic}

The initial conditions are very important to treat consistently.
Due to the rapid variation of $X(a)\sim a^{-6/(2n-1)}$ from
Eq.~(\ref{eq:xearly}), or equivalently $x_\phi\approx -\wde$,
it is not practical to use too small an
initial scale factor $a_i$. Indeed, since
\be
\phi=\int d\phi=\int d\ln a \,\frac{\dot\phi}{H}\ , \label{eq:phiint}
\ee
then $\phi\sim a^{3(1-2\wde)/2}$.
As the initial evolution is locked in by
the charge conservation behavior, the behavior is  insensitive
to $a_i$. To prevent issues with numerical instability we
adopt $a_i=10^{-2}$, in the matter dominated epoch.

The practical steps are
(all variables below indicate their initial values only,
and we indicate values for our fiducial $n=1/4$):
\begin{enumerate}
\item Set $\Omega_{{\rm de},i}\approx a_i^{-3/(2n-1)}=a_i^6$.
We vary this in order to
obtain the desired $\Omega_{{\rm de},0}$.
\item Next, we set $x_{g,i}=\alb=(3/5)\ode$. This comes from
Eq.~(\ref{eq:albss}) and Eq.~(\ref{eq:fearly}) with
$f_i=f_{\rm early}=3/5$ ($f_i$ will lie in a different range
for other $n$).
\item
Using Eq.~(\ref{eq:kinfrac}), $x_{k,i}$ is given by
\be
x_{k,i}=\frac{1-f_i}{2n-1}\,\Omega_{{\rm de},i} .
\ee
This will also work with $\alk$ to
guarantee a ghost free initial condition.
\item From Eq.~(\ref{eq:xearly}) we get $x_{\phi,i}=-1/(2n-1)=2$.
\item Set $x_{f,i}=3x_{\phi,i}+3/2=15/2$. This arises from
Eq.~(\ref{eq:phiint}) in the matter dominated era.
\item Finally we set $x_{\lam,i}=\Omega_{{\rm de},i}^{7/4}$. This
comes from Eq.~(\ref{eq:xlamp}) and the early time scaling
of $\ode(a)$. We can vary this in order to obtain different late
time behavior. Note that since
$\Omega_{{\rm de},i}^{7/4}\approx a_i^{21/2}$ the numerical
precision at $a_i=10^{-2}$ must be treated carefully.
\end{enumerate}
Values of the quantities at successive timesteps are determined by the evolution equations (for $\ode$, $x_k$, $x_g$, $x_\lam$, $x_f$) and constraint equations (for $x_\phi$, $\wde$, $\wtot$).

The integration scheme of the numerical dynamical system of equations is implemented as an interactive notebook \github{https://rcalderonb6.github.io/phantom-X/} allowing further exploration for user-specified parameter values.

\end{document}